\documentclass[wssquare]{ws-procs975x65} 
\begin{document}
\title{Antiferromagnetism with Ultracold Atoms}

\author{Randall G. Hulet$^*$, Pedro M. Duarte, Russell A. Hart and Tsung-Lin Yang}

\address{Department of Physics and Astronomy, Rice University,\\
Houston, Texas 77005, USA\\
$^*$E-mail: randy@rice.edu\\
atomcool.rice.edu}

\begin{abstract}

We use ultracold spin--1/2 atomic fermions ($^6$Li) to realize the Hubbard model on a three-dimensional (3D) optical lattice. At relatively high temperatures and at densities near half-filling, we show that the gas forms a Mott insulator with unordered spins.  To observe antiferromagnetic order that is predicted to occur at lower temperatures, we developed the compensated optical lattice method to evaporatively cool atoms in the lattice.  This cooling has enabled the detection of short-range magnetic order by spin-sensitive Bragg scattering of light.
\end{abstract}


\bodymatter

\section{Introduction}\label{intro}
Cold atomic gases have emerged as a versatile new platform for applications to fundamental many-body physics \cite{Bloch08}.  This versatility arises from a remarkable ability to precisely control nearly every relevant system parameter.  Most notably, the density, interaction, the presence or complete absence of disorder, and dimensionality can all be controlled.  The one glaring, and severely limiting exception to this list, is temperature.  While the absolute temperatures of cold gases are lower than for any other man-made or naturally occurring physical system, the absolute temperature is not the most important quantity.  Rather, the relevant energy scale is set by the Fermi temperature $T_F$, which determines the absolute temperature $T$ where phase transitions to superfluid or quantum magnetic states occur. (For bosons, the analogous energy scale is the Bose-Einstein condensation temperature, which is nearly identical to $T_F$).  The Fermi energy in cold gases is constrained to be about $1\,\mu$K by the need to keep densities low enough to minimize molecular recombination.  In contrast, the Fermi temperature of a high-temperature superconductor, at $\sim$10,000 K, is nearly $10^{10}$ times higher.  In these materials, a transition temperature as high as 100 K is only 1\% of $T_F$, well below the relative temperatures achieved in state-of-the-art atomic fermion experiments, for which the most advanced cooling methods, performed by evaporative cooling of atoms in traps, have produced $T / T_F \simeq 0.04$.

For cold atomic gases to fully attain their potential to realize and characterize new quantum states of matter requires that we develop methods for cooling to ever lower temperatures.  In our work, we are particularly motivated to improve our understanding of high-temperature superconductors, where the ultimate goal is to create new materials with even higher transition temperatures than currently possible.

Our specific interest, and an ideal test-bed for our cooling methods, is the Hubbard model, which was originally proposed to describe the magnetic and conduction properties of electrons in transition metal oxides.  Shortly after the discovery of high-temperature superconductors, Anderson suggested that the Hubbard model may also contain the essential ingredients to describe their remarkable properties \cite{Anderson87}.  The Hubbard model has now become one of the most important models of strongly correlated matter, and was recognized early-on as a target for emulation with ultracold atoms on an optical lattice \cite{Hofstetter02,Jaksch05}.  The model itself is deceptively simple, describing a spin-1/2 Fermi gas on a lattice by its hopping rate \textit{t}, and an on-site interaction energy \textit{U}. Only the ground band need be considered when both \textit{T} and \textit{U} are much less than the band gap.  In this case, the Pauli principle forbids atoms of the same spin from occupying the same lattice site.

Despite its simplicity, the solutions of the Hubbard Hamiltonian are remarkably complex depending on the sign and magnitude of the ratio $U/t$, as well as the density \textit{n}.  Similarly, the phase  diagram of a typical high-$T_c$ material exhibits a variety of complex behaviors, including antiferromagnetism, a poorly understood pseudo-gap region in which there are pair correlations, but no long-range order, and at very low temperatures, \textit{d}-wave superconductivity.  It is unknown whether the Hubbard model is sufficient to provide a unifying theory of high-$T_c$ materials, because it cannot be solved with analytic or numerical methods since the basis size, which scales as $2^N$, where \textit{N} is the total number of particles, quickly overwhelms the computational capability of any current or envisioned digital computer.  By using cold atoms on a lattice, we create an analog quantum computer specifically designed to solve the Hubbard model.

\section{Experiment}
The experiment utilizes the $|F=1/2; m_{F}=+1/2\rangle$ and $|F=1/2;
m_{F}=-1/2\rangle$ hyperfine states of $^{6}$Li, which we label $|\uparrow\rangle$ and
$|\downarrow\rangle$, respectively. Our methods have been described
previously~\cite{Duarte15,Hart15}.  We use all-optical methods \cite{Duarte11} to confine and cool the atoms, before evaporatively cooling them in a crossed-beam optical trap.  These atoms are then loaded into
a simple cubic optical lattice. The temperature of the atoms in the trap, prior to loading them into the lattice, is measured to be $T/T_{F} = 0.04\pm0.02$ by fitting the density distribution after time of flight.

The optical lattice is formed by three retroreflected red-detuned ($\lambda = 1064$ nm)
Gaussian laser beams of depth $V_{0}=7\,E_{r}$, where $E_{r} = \frac{ \hbar^{2} \pi^{2}}{ 2ma^{2} } = 1.4\,\mu$K is the recoil energy, $m$ is the mass of the atoms, and $a = \lambda/2$ is the lattice spacing.  We use the broad Feshbach resonance in $^{6}$Li at 832~G~\cite{Houbiers98,Zurn13} to set the on-site interaction strength, $U$.  Typical experimental parameters are $V_{0}=7\,E_{r}$, $t = 0.038\,E_{r}$, and $U = 0.38\,E_{r}$, corresponding to an \textit{s}-wave scattering length of $260\,a_0$.

\begin{figure}
\begin{center}
\includegraphics[width=4.8in]{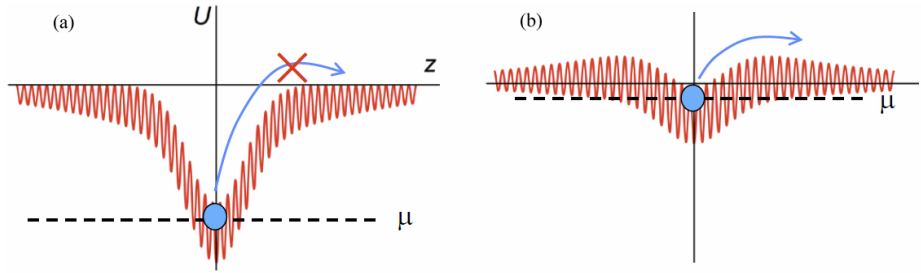}
\end{center}
\caption{(a) Optical lattice with the usual Gaussian confinement, which inhibits evaporation;  (b)  Compensated optical lattice showing evaporation.}
\label{RGH:Fig1}
\end{figure}

\begin{figure}[b]
\begin{center}
\includegraphics[width=4.8in]{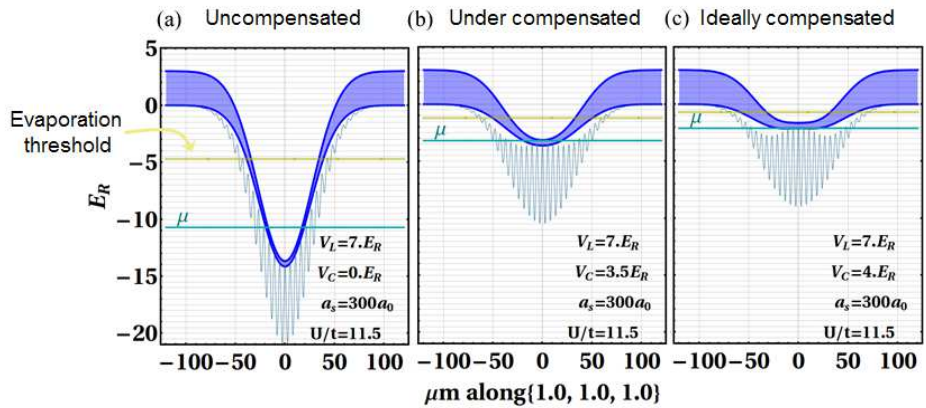}
\end{center}
\caption{(a) Uncompensated optical lattice; (b) Under compensated lattice; and (c) Ideally compensated lattice.  The oscillating light blue lines represents the lattice, where the lattice constant is exaggerated for clarity.  The dark blue region is the lowest Bloch band.  Compensation can both bring $\mu$ close to the evaporation threshold, and with the proper relative beam sizes, it can also flatten the band, as shown in (c). In all three panels $\mu$ corresponds to a central density of $n=1$.}
\label{RGH:Fig2}
\end{figure}

\subsection{Compensated Optical Lattice}
Temperatures of fermions in traps can usually be cooled to a factor of $\sim$3-4 below that in optical lattices.  There are several factors that contribute to this situation, including technical noise sources in lattices that are more pernicious than in traps, but the primary reason is that the depth of a trap may be adjusted such that the chemical potential, $\mu$, lies just below the lip of the trap. A small, but steady rate of evaporation mitigates heating to produce relative temperatures of $T/T_{F} \simeq 0.04$.  In a standard lattice, however, the Gaussian beams create a confining envelope that cannot be adjusted without also affecting the lattice depth, $V_{0}$.  Since $\mu$ lies far below the lip of the confinement potential at densities of $n = 1$ (one atom per lattice site or ``half-filling"), evaporation is impeded by the small Boltzmann factor, as shown in Fig.~\ref{RGH:Fig1}(a).  We introduce additional degrees of freedom, provided by blue-detuned (532 nm) anti-confining compensation beams, to control the depth of confinement independently of $V_{0}$.  These compensation beams overlap each of the lattice beams but are not themselves retroreflected~\cite{Mathy12,Hart15}. This provides the ability to tune the depth of confinement such that atoms may evaporate, as shown in Fig.~\ref{RGH:Fig1}(b). With the additional freedom to choose the Gaussian waists of the compensating beams relative to the lattice beams, it is also possible to flatten the lowest band, and thereby increase the volume fraction where $\mu$ is nearly constant \cite{Mathy12}. The effect of compensation on the band structure is illustrated in Fig.~\ref{RGH:Fig2}.

\begin{figure}[b]
\begin{center}
\includegraphics[width=2.3in]{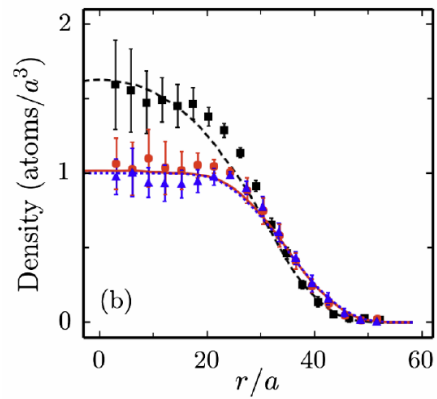}
\end{center}
\caption{Density distributions showing the development of the Mott plateau at $n=1$ with increasing $U/t$.  These 3D distributions are obtained from the column density images by an Abel transform.  The black squares, red dots, and blue triangles correspond to $U/t= 3.1$, $11.1$, and $14.5$, respectively. (Reprinted from Ref.~\citenum{Duarte15}). }
\label{RGH:Fig3}
\end{figure}

\section{Results}
\subsection{Mott Insulator}
The Hubbard model for densities near $n=1$ and for $T\ll U$ describes a system which undergoes a smooth crossover to a Mott insulating regime,
characterized by a suppression of density fluctuations and a reduction of the number of doubly occupied sites.  The suppression of density fluctuations implies a reduction of the
compressibility~\cite{DeLeo08}.  Figure \ref{RGH:Fig3} shows the resulting plateaus in the measured trapped density distributions at $n=1$ when $U/T$ becomes sufficiently large.  We have used these density distributions to extract the local compressibility of the trapped gas as a function of \textit{n} and have shown that this quantity can provide accurate thermometry in the lattice down to temperatures $T \simeq t$.  The Mott transition in the Fermi-Hubbard model has been studied in previous experiments by measuring the variation of the bulk double occupancy with atom number~\cite{Jordens08} and the
response of the cloud radius to changes in external confinement~\cite{Schneider08}, both of which are related to the global compressibility, and are severely suppressed for large interactions.

\begin{figure}[t]
\begin{center}
\includegraphics[width=4.5in]{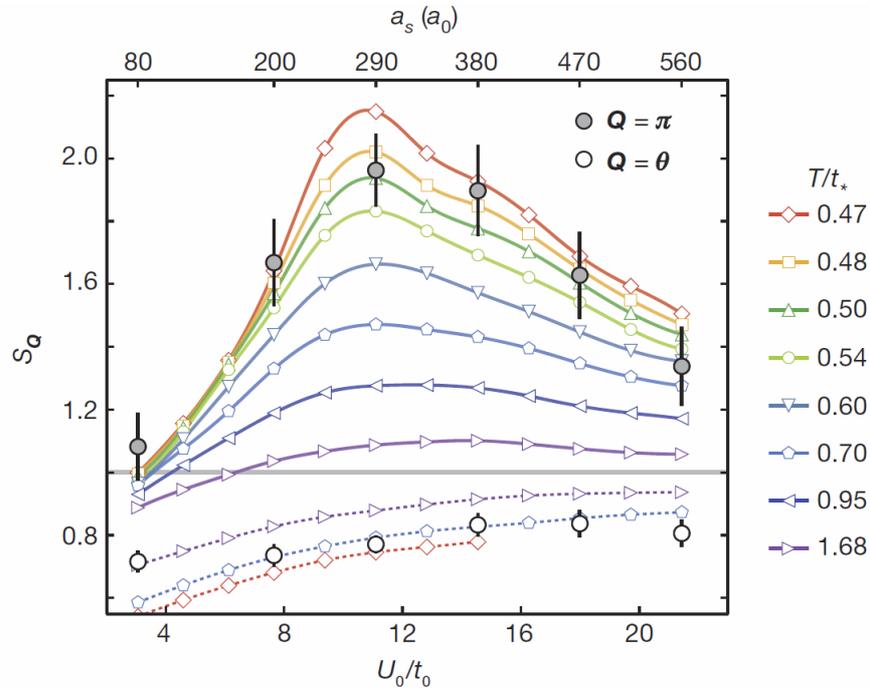}
\end{center}
\caption{Bragg scattering signal.  Data shown by gray dots corresponds to the spin-structure factor in the coherently scattered direction (satisfying the Bragg condition with a reciprocal lattice vector $Q = (2\pi/a)(\frac{1}{2},\frac{1}{2},\frac{1}{2}))$. The different symbols and colored lines are the results of the QMC calculations for the different temperatures indicated.  Data given by open circles correspond to a non-Bragg angle, $\theta$.  $S_\theta < 1$ because doubly-occupied sites scatter no light and there is no coherent enhancement in this direction. $S_Q = 1$ in the high-temperature limit for any direction.   (Reprinted from Ref.~\citenum{Hart15}). }
\label{RGH:Fig4}
\end{figure}

\subsection{Antiferromagnetic Order}
At very low temperatures, the Mott insulator undergoes a phase transition to an antiferromagnetic (AFM) state.  This transition occurs at the N\'{e}el temperature, $T_{N}$ ($\sim\!4t^{2}/U$ for $U\gg t$).  $T_N$ is more than a factor of two below the lowest temperatures previously attained in a isotropic 3D lattice \cite{Greif13,Imriska14}.  In our experiment, we use Bragg scattering of near-resonant light \cite{Corcovilos10} to detect the onset of AFM order \cite{Hart15}. Detection of magnetic order requires that the light scattering be spin-dependent.  We accomplished this by tuning the laser between the transition frequencies of the two states, $|\uparrow\rangle$ and $|\downarrow\rangle$, of our spin-1/2 system.

The strength of the Bragg signal, which is directly proportional to the spin structure factor, $S_\pi$, is shown as the solid dots in Fig.~\ref{RGH:Fig4} as a function of $U/t$ \cite{Hart15}.  As expected from quantum Monte Carlo (QMC) calculations, $S_\pi$ exhibits a broad maximum centered near $U/t \approx 10$ \cite{Paiva11}.  The results of QMC for various temperatures are also shown in Fig.~\ref{RGH:Fig4}.  The QMC calculations are performed for a homogeneous density distribution and then averaged over the measured density distribution using the local density approximation.  The calculations demonstrate that the Bragg signal is extremely sensitive to temperature in this regime, and thus the combination of Bragg scattering and QMC provides sensitive thermometry in a regime where previously there was none.  Comparison of the data with QMC indicates that the temperature is $T/t =0.51\pm0.06$, where the uncertainty is due to the statistical error in the measured $S_\pi$.  In terms of $T_{N}$, the temperature is $T/T_{N}=1.42\pm0.16$.

\section{Conclusions and Outlook}
In conclusion, we have detected short-range antiferromagnetic correlations in the Hubbard model.  At 1.4 $T_{N}$, the correlations are still short-range.  We are able to observe these correlations because the compensated optical lattice produces temperatures more than a factor of 2 lower than previously attained for fermions in a 3D optical lattice, and because Bragg scattering of light is an extremely sensitive detector.  We must cool further in order to get below $T_N$, and more generally, to open up the study of novel quantum states of matter that have yet to be created.  We are optimistic that the compensated lattice method can be pushed further to optimize evaporative cooling in the lattice and to flatten the ground band more completely.  To do so, we are using a spatial light modulator to produce compensation intensity distributions that are more complex than just the simple Gaussian beams employed here.  In addition, we are exploring ways to use blue-detuned light to imprint metallic regions with $n < 1$ on the Mott plateau as a way of storing entropy or conducting it away from the low-entropy, but insulating Mott phase.

\section*{Acknowledgments}
We gratefully acknowledge the invaluable contributions by our theory collaborators T. Paiva, E. Khatami, R. T. Scalettar, N. Trivedi, and D. A. Huse. This work was supported by NSF grant No.\,PHY-1408309, the ONR, the Welch Foundation (Grant No. C-1133), and an ARO-MURI grant No.\,W911NF-14-1-003.

\bibliographystyle{ws-procs975x65}
\bibliography{Hulet_ICOLS_2015}

\end{document}